\documentstyle[12pt]{article}
\input{epsfig.sty}

\newcommand{\be}{\begin{eqnarray}}
\newcommand{\ee}{\end{eqnarray}}

\def\lsim{\mathrel{\rlap{\lower4pt\hbox{\hskip1pt$\sim$}}
    \raise1pt\hbox{$<$}}}               % less than or approx. symbol
\def\gsim{\mathrel{\rlap{\lower4pt\hbox{\hskip1pt$\sim$}}
    \raise1pt\hbox{$>$}}}               % greater than or approx. symbol

\begin{document}

\vspace{0.5cm}

\rightline{Preprint INFN-GE 97/1}
\rightline{INFN-ISS 97/1}

\vspace{1cm}

\begin{center}

\LARGE{Bloom-Gilman duality of inelastic structure functions in nucleon
and nuclei\footnote{\bf To appear in Physical Review C.}}\\

\vspace{1.5cm}

\large{G. Ricco$^{(*, **)}$, M. Anghinolfi$^{(**)}$, M.
Ripani$^{(**)}$,\\ S. Simula$^{(***)}$ and M. Taiuti$^{(**)}$}\\

\vspace{0.5cm}

\normalsize{$^{(*)}$Dipartimento di Fisica, Universit\`a di Genova\\ Via
Dodecanneso 33, I-16146, Genova, Italy\\$^{(**)}$Istituto Nazionale di
Fisica Nucleare, Sezione di Genova\\ Via Dodecanneso 33, I-16146,
Genova, Italy\\$^{(***)}$Istituto Nazionale di Fisica Nucleare, Sezione
Sanit\`{a}\\ Viale Regina Elena 299, I-00161 Roma, Italy}\\

\end{center}

\vspace{1cm}

\begin{abstract}

\noindent The Bloom-Gilman local duality of the inelastic structure
function of the proton, the deuteron and light complex nuclei is
investigated using available experimental data in the squared
four-momentum transfer range from $0.3$ to $5 ~ (GeV/c)^2$. The results
of our analysis suggest that the onset of the Bloom-Gilman local duality
is anticipated in complex nuclei with respect to the case of the proton
and the deuteron. A possible interpretation of this result in terms of a
rescaling effect is discussed with particular emphasis to the
possibility of reproducing  the damping of the nucleon-resonance
transitions observed in recent electroproduction data off nuclei.

\end{abstract}

\vspace{1cm}


\newpage

\pagestyle{plain}

\section{Introduction.}

\indent The investigation of inelastic lepton scattering off nucleon and
nuclei can provide relevant information on the concept of parton-hadron
duality, which deals with the relation among the physics in the
nucleon-resonance and Deep Inelastic Scattering ($DIS$) regions. As is
known, well before the advent of $QCD$, parton-hadron local duality was
observed empirically by Bloom and Gilman \cite{BG70} in the structure
function $\nu W_2^p(x, Q^2)$ of the proton measured at $SLAC$ (where $x
\equiv Q^2 / 2m \nu$ is the Bjorken scaling variable, $m$ the nucleon
mass and $Q^2$ the squared four-momentum transfer). Moreover, it is well
established that both the electroexcitation of the most prominent
nucleon resonances and the nuclear structure function in the $DIS$
region are affected by nuclear medium (cf., e.g., Refs.
\cite{LA86,AR94}). In particular, existing data on the electroproduction
of nucleon resonances show that the disappearing of the resonance bumps
with increasing $Q^2$ is faster in nuclei than in the nucleon (cf.,
e.g., Refs. \cite{HE74}, \cite{GE96} and references therein quoted).
Therefore, in this paper we want to address the specific question
whether and to what extent the Bloom-Gilman duality already observed in
the proton occurs also in the structure function of a nucleus. To this
end, all the available experimental data for the structure functions of
the proton, the deuteron and light complex nuclei in the $Q^2$ range
from $0.3$ to $5 ~ (GeV/c)^2$ have been analyzed and the $Q^2$
behaviours of the structure function and its moments are presented for
all the targets considered. In case of the proton we observe that the
Bloom-Gilman local duality is fulfilled only by the inelastic part of the
structure function, while the inclusion of the contribution of the
elastic peak leads to remarkable violations of the local duality. In case
of complex nuclei, despite the poor statistics of the available data, it
is found that the onset of the parton-hadron local duality for the
inelastic part of the structure function is anticipated with respect to
the case of the proton and the deuteron. Nevertheless, new high-precision
nuclear data are needed and, in this respect, it should  be mentioned
that the Thomas Jefferson National Accelerator Facility ($TJNAF$) is
expected to provide in the next future systematic measurements  with
unprecedented accuracy of the nucleon and nuclear inelastic response to
electron probes in the region of nucleon-resonance production for values
of $Q^2$ up to several $(GeV/c)^2$. Finally, a possible interpretation
of the observed nuclear modification of the onset of the Bloom-Gilman
local duality in terms of a $Q^2$-rescaling effect is discussed with
particular emphasis to the possibility of reproducing the damping of the
nucleon-resonance transitions observed in recent electroproduction data
off nuclei \cite{GE96}. 

\section{The {\em dual} structure function.}

\indent The Bloom-Gilman local duality \cite{BG70} states that the
smooth scaling curve measured in the $DIS$ region at high $Q^2$
represents an average over the resonance bumps seen in the same $x$
region at low $Q^2$. More precisely, Bloom and Gilman pointed out the
occurrence of a precocious scaling of the average of the inclusive $\nu
W_2^p(x', Q^2)$ data in the resonance region to the $DIS$ structure
function $F_2^p(x')$, at corresponding values of an improved empirical
variable $x' = x / (1 + m^2 x / Q^2)$. Later on, within $QCD$, a
justification of the Bloom-Gilman duality was offered by De Rujula,
Georgi and Politzer \cite{RGP77} in terms of the moments $M_n(Q^2)$ of
the nucleon structure function $F_2(\xi, Q^2)$:
 \be
    M_n(Q^2) \equiv \int_0^1 d\xi \xi^{n-2} F_2(\xi, Q^2)
    \label{1}
 \ee
where $\xi$ is the Nachtmann variable (cf. \cite{GP76})
 \be
    \xi = {2x \over 1 + \sqrt{1 + {4 m^2 x^2 \over Q^2}}}
    \label{2}
 \ee 
Using the Operator Product Expansion ($OPE$) the authors of Ref.
\cite{RGP77} argued that
 \be
    M_n(Q^2) = A_n(Q^2) + \sum_{k=1}^{\infty} \left ( n {\gamma^2
    \over Q^2} \right )^k B_{nk}(Q^2)
    \label{3}
 \ee
where $\gamma^2$ is a scale constant. The first term $A_n(Q^2)$ in Eq.
(\ref{3}) is the result of perturbative $QCD$, while the remaining
terms $B_{nk}(Q^2)$ are higher twists related to parton-parton
correlations. The value of $\gamma^2$ is relatively small (a recent
estimate, made in Ref. \cite{JU95}, yields $\gamma^2 \sim 0.1 \div 0.3
~ GeV^2$). Therefore, at $Q^2 \gsim m^2$ the asymptotic moments
$A_n(Q^2)$ are still leading, while resonances contribute to the higher
twists $B_{nk}(Q^2)$. The quantities $A_n(Q^2)$ can be considered as
the moments of a smooth structure function, which can be identified with
the average function $\langle \nu W_2(x, Q^2) \rangle$ occurring in
the Bloom-Gilman local duality, namely \cite{GP76}:
 \be
    \langle \nu W_2 (x, Q^2) \rangle & = & {x^2 \over (1 + {4 m^2 
    x^2 \over Q^2})^{3/2}} {F_2^S(\xi, Q^2) \over \xi^2}
    \nonumber \\
    & + & 6 {m^2 \over Q^2} {x^3 \over (1 + {4 m^2 x^2
    \over Q^2})^2} \int_{\xi}^1 d\xi' {F_2^S(\xi', Q^2) \over \xi'^2}
    \nonumber \\
    & + & 12 {m^4 \over Q^4} {x^4 \over (1 + {4 m^2 x^2
    \over Q^2})^{5/2}} \int_{\xi}^1 d\xi' \int_{\xi'}^1 d\xi" 
    {F_2^S(\xi", Q^2) \over \xi"^2}
    \label{4}
 \ee
where the $\xi$-dependence as well as the various integrals appearing in
the r.h.s. account for target mass effects in the $OPE$ of the hadronic
tensor. According to Ref. \cite{GP76} these effects have to be included
in order to cover the low $Q^2$ region. In Eq. (\ref{4}) $F_2^S(x, Q^2)$
represents the asymptotic nucleon structure function, fitted to high
$Q^2$ proton and deuteron data \cite{AR95} and extrapolated down to low
values of $Q^2$ by the Altarelli-Parisi evolution equations \cite{AP77}.
In this paper the Gluck-Reya-Vogt ($GRV$) fit \cite{GRV92}, which
assumes a renormalization scale as low as $0.4 ~ (GeV/c)^2$, will be
used to obtain the parton densities $\rho_f(x, Q^2)$ evolved at Next
to Leading Order ($NLO$) from sufficiently low $Q^2$ to cover the range
of interest in the present analysis. In the $DIS$ region \cite{AEM78}
one gets $F_2^S(x, Q^2) =$ $\sum_f e_f^2 x \left [ \rho_f(x, Q^2) +
\bar{\rho}_f(x, Q^2) \right ]$.

\indent In what follows, we will refer to the mass-corrected, $NLO$
evolved function (\ref{4}) as the {\em dual} structure function of the
nucleon. We stress that by definition the {\em dual} function does not
contain any higher twists generated by parton-parton correlations, i.e.
the twists related to the moments $B_{nk}(Q^2)$ in Eq. (\ref{3}). It
should be mentioned that Eq. (\ref{4}) suffers from a well known
\cite{MMG89,BJT79} mismatch; indeed, since $\xi(x =1) = 2 / (1 +
\sqrt{1 + 4 m^2 / Q^2}) < 1$, the r.h.s. of Eq. (\ref{4}) remains
positive, while its l.h.s. vanishes, as $x$ approaches the elastic
end-point $x = 1$. An alternative approach \cite{MMG89}, limited at the
twist-4 order, is well behaved at the kinematical $x = 1$ threshold,
but it cannot be extrapolated to low values of $Q^2$, because an
expansion over the quantity $m^2 x^2 / Q^2$ is involved. Moreover,
inelastic threshold effects, due to finite pion mass, are accounted for
neither in the $Q^2$ evolution nor in the target mass corrections,
because they are basically higher-twist effects; then, in order to
make a detailed comparison with experimental data in the low $Q^2$
region, we set the {\em dual} structure function (\ref{4}) to zero at $x
\ge x_{th}$, where
 \be
    x_{th} = {1 \over 1 + {m_{\pi}^2 + 2 m_{\pi} m \over Q^2}}
    \label{5}
 \ee
with $m_{\pi}$ being the pion mass. Therefore, the investigation of
parton-hadron local duality will be limited to values of $x$ not
larger than $x_{th}$ (\ref{5}) and to a $Q^2$ range between a minimum
value $Q_{min}^2$, which is of the order of the mass scale $\mu^2$
where the moments of the structure functions $M_n(Q^2)$ start to evolve
according to twist-2 operators, and a maximum value $Q_{max}^2$ ($\gsim
5 ~ (GeV/c)^2$ \cite{JU95}), where the resonance contribution to the
lowest moments of Eq. (\ref{3}) is of the same order of magnitude of the
experimental errors.

\section{Inclusive data analysis.}

\indent The pseudo-data for the proton structure function $\nu W_2^p$,
obtained from a fit of $SLAC$ data at medium and large $x$ \cite{BO79}
and from a fit of the $NMC$ data \cite{AR95} in the $DIS$ region, are
reported in Fig. 1 versus the Natchmann variable $\xi$ (\ref{2}) and
compared with the {\em dual} structure function (\ref{4}) at fixed
values of $Q^2$ in the range $0.5$ to $2 ~ (GeV/c)^2$. As already
pointed out in \cite{RGP77}, the onset of local duality occurs at $Q^2
\simeq Q_0^2 \sim 1 \div 2 ~ (GeV/c)^2$.

\indent In case of the deuteron the average nucleon structure function
$\nu \bar{W}_2^N \equiv (\nu W_2^p + \nu W_2^n) / 2$, obtained from
(\ref{4}), should be folded with the momentum distribution which
accounts for the internal motion of the nucleon in the deuteron. The
most evident effect of this folding is a broadening of the nucleon
elastic peak occurring at $x = 1$ into a wider quasi-elastic peak,
partially overlapping the inelastic cross section at large $x$. In the
deuteron the nucleon momentum distribution is relatively narrow and
this fact limits the overlap to the kinematical regions corresponding
to $x \gsim 0.8$. Therefore, it is still possible to subtract the
quasi-elastic contribution directly from the total cross section
\cite{BO79}. The folding of the {\em dual} structure function of the
nucleon (Eq. (\ref{4})) with the nucleon momentum distribution in the
deuteron is performed following the procedure of Ref. \cite{SIM95},
which can be applied both at low and high values of $Q^2$, at variance
with the standard high-$Q^2$ folding of Ref. \cite{FS81}. Adopting the
deuteron wave function corresponding to the Paris nucleon-nucleon
potential \cite{PARIS}, our results for the {\em dual} structure
function folded in the deuteron are reported in Fig. 2 at fixed values
of $Q^2$ in the range from $0.5$ to $2 ~ (GeV/c)^2$, and compared with
the deuteron structure function per nucleon, $F_2^D(\xi, Q^2) \equiv
\nu W_2^D(\xi, Q^2) / 2$, obtained from the $NMC$ data \cite{AR90} at
low $\xi$ and the fits of inclusive data given in Refs. \cite{AR95} and
\cite{BO79}. Despite the smoothing of nucleon resonances caused by the
Fermi motion, parton-hadron local duality appears to start again at $Q^2
\simeq Q_0^2 \sim 1 \div 2 ~ (GeV/c)^2$.

\indent In case of complex nuclei the analysis of existing inclusive
data is complicated by different reasons:
 
\begin{itemize}

\item inclusive data, coming mainly from old experiments generally with
poor statistics, is still very fragmented; 

\item the longitudinal to transverse separation has been done only in
case of few measurements carried out in the nucleon-resonance region. An
experiment performed at $Q^2 = 0.1 ~ (GeV/c)^2$ \cite{BA88} in $^{12}C$
and $^{56}Fe$ claims a longitudinal to transverse ratio compatible with
zero inside experimental errors ($\sim 10\%$), while at $Q^2 \gsim 1 ~
(GeV/c)^2$ deep inelastic scattering data are consistent with larger,
fairly $x$ independent $\sigma_L / \sigma_T$ ratios \cite{RO97}.
However, the sensitivity of the extraction of the nuclear structure
function $\nu W_2^A$ from the total cross section to the longitudinal
to transverse ratio turns out to be rather small (not larger than few
$\%$ when $\sigma_L / \sigma_T$ is moved from zero to $20 \%$ in the
worst kinematical conditions). Therefore, we have simply interpolated
all the existing data for $\sigma_L / \sigma_T$ in proton and nuclei,
averaged on $x$, as a function of $Q^2$ only, obtaining the following
empirical ratio:
 \be
    \sigma_L / \sigma_T = a \cdot Q^2 \left [ e^{-b Q^2} + c \cdot e^{-d
    Q^2} \right ] \nonumber
    \label{6}
 \ee
with $a = 0.014$, $b = 0.07$, $c = 40.8$ and $d = 0.78$. Since the
observed dependence of the inclusive nuclear data on the mass number $A$
is weak \cite{GO94}, the nuclear structure function $\nu W_2^A$ has been
determined as a function of $\xi$ for fixed $Q^2$ bins using data for
$^9Be$ \cite{GO94,FR81}, $^{12}C$ \cite{BO79,AR90,DA93} and $^{16}O$
\cite{GE96}. Therefore, our result could be considered representative of
a complex nucleus with $A \simeq 12$;

\item since the nucleon momentum distribution is wider in complex
nuclei than in the deuteron (cf. \cite{CS96} and references therein
quoted), the quasi-elastic contribution strongly overlaps the inelastic
cross section at low values of $Q^2$; moreover, the quasi-elastic
peak is known to be affected by final state interaction effects at
$Q^2 \lsim 0.5 ~ (GeV/c)^2$ (cf. \cite{GE96}). Then, the subtraction of
the quasi-elastic contribution is more critical in nuclei;

\item nucleon binding can affect nuclear structure function and should
be properly taken into account \cite{AR94};

\end{itemize}

\indent The calculation of the quasi-elastic contribution has been
performed following the approach of Ref. \cite{CS96}, which has been
positively checked against $SLAC$ data at values of $Q^2$ of few
$(GeV/c)^2$ \cite{CS96} as well as against jet-target data at lower
$Q^2$ (down to $0.1 ~ (GeV/c)^2$ \cite{GE96}). An example of the
quality of the agreement among the parameter-free predictions of the
quasi-elastic contribution to the inclusive $^{12}C(e, e')X$ cross
section and available data at the quasi-elastic peak and in its
low-energy side for a $Q^2$ range from $0.5$ to $1 ~ (GeV/c)^2$, is
shown in Fig. 3. Furthermore, the {\em dual} structure function of the
nucleon (\ref{4}) has been folded using again the procedure of Ref.
\cite{SIM95}, which, in case of complex nuclei, involves the nucleon
spectral function of Ref. \cite{CS96}; in this way binding effects are
taken into account for states both below and above the Fermi level.

\indent After quasi-elastic subtraction, the results obtained for the
inelastic nuclear structure function per nucleon, $F_2^A(\xi, Q^2)
\equiv \nu W_2^A(\xi, Q^2) / A$, are shown in Fig. 4 for various $Q^2$
bins, namely: $Q^2 = 0.375 \pm 0.03$; $0.50 \pm 0.05$; $0.75 \pm 0.05$;
$1.1 \pm 0.2$; $1.8 \pm 0.2$ and $4.5 \pm 0.5 ~ (GeV/c)^2$. In
comparison with the proton and deuteron cases, the most striking
feature of our result for nuclei with $A \simeq 12$ is a more rapid
smoothing of the resonance bumps with increasing $Q^2$, which favours a
faster convergence towards the {\em dual} structure function in the
nucleus. This effect, already noticed in Ref. \cite{GE96} in the $Q^2$
range from $0.1$ to $0.5 ~ (GeV/c)^2$, cannot be completely explained
as a broadening of the resonance width due to final state interactions,
but some extra-damping factor is needed in order to reproduce the
missing resonance strength\footnote{Medium effects were noticed also
in Ref. \cite{CE94} in the form of a better $y$-scaling of the
inclusive cross section in the region of $P_{33}(1232)$ resonance
electroproduction in nuclei with respect to the free nucleon case.}.
From Fig. 4 it can clearly be seen that the {\em dual} structure
function (\ref{4}), properly folded \cite{SIM95} using the nucleon
spectral function of Ref. \cite{CS96}, approaches the inelastic data at
$Q^2 \simeq Q_0^2 \sim 0.5 \div 1.1 ~ (GeV/c)^2$, i.e. for values of
$Q_0^2$ lower than those found in the proton and in the deuteron
($Q_0^2 \sim 1 \div 2 ~ (GeV/c)^2$). 

\section{Results and discussion.}

All the data obtained for the structure function $F_2^p(\xi, Q^2)$ in
the proton, $F_2^D(\xi, Q^2)$ in the deuteron and $F_2^A(\xi, Q^2)$ in
nuclei with $A \simeq 12$ (see Figs. 1, 2 and 4) show a systematic
approach to the properly folded dual structure function for $\xi \gsim
0.2$, the convergence being even more evident and faster in nuclei, due
to the fading of resonances already at $Q^2 \gsim 0.5 ~ (GeV/c)^2$.

\indent At smaller values of $\xi$ and for $Q^2 < 5 ~ (GeV/c)^2$ proton
and deuteron data seems to be compatible with an evolution of sea
partons slower than the $GRV$ prediction; in complex nuclei the
difference is further enhanced by the shadowing effect \cite{AR94}. The
reason of this discrepancy is not evident and various motivations can
be invoked, like: a breakdown of duality and/or (unexpected) higher
twists at low $x$, or, more likely, a not yet completely consistent
initial ($Q^2 \simeq 0.4 ~ (GeV/c)^2$) parton density parametrization in
$GRV$ fit \cite{GRV92}.

\indent The minimum momentum transfer $Q_0^2$, where local duality
provides an acceptable fit to the average inclusive inelastic cross
sections, should be related to the mass scale $\mu^2$, which is defined
as the minimum value of $Q^2$ where the moments of the structure
functions $M_n(Q^2)$ begin to evolve according to twist-2 operators, i.e.
when $M_n(Q^2) \sim A_n(Q^2)$. Thus, the $Q^2$-dependence of all the
moments should exhibit a systematic change of the slope at $Q^2 \simeq
Q_0^2 \simeq \mu^2$ and then should follow the perturbative $QCD$
evolution. The {\em experimental} $M_2(Q^2)$, $M_4(Q^2)$, and $M_6(Q^2)$
moments have been computed for the proton, the deuteron and nuclei with
$A \simeq 12$ according to Eq. (\ref{1}) using the experimental data for
the structure function $F_2^p(\xi, Q^2)$, $F_2^D(\xi, Q^2)$ and
$F_2^A(\xi, Q^2)$ shown in Figs. 1, 2 and 4. The results are plotted in
Fig. 5 for $Q^2$ between $0.3$ and $5 ~ (GeV/c)^2$. The expected
systematic change of the slope is clearly exhibited by all the moments
considered and the corresponding values of $Q^2 = Q_0^2$ are reported in
Table 1. Moreover, in Fig. 6 the relative deviation
 \be
    {\Delta M_n \over A_n}(Q^2) \equiv {M_n(Q^2) - A_n(Q^2) \over
    A_n(Q^2)}
    \label{7}
 \ee
is plotted as a function of $Q^2$, where $A_n(Q^2)$ has been computed
from the properly folded {\em dual} structure function (continuous
lines in Figs. 1, 2 and 4). Despite the large errors affecting the
available nuclear data, in case of all the targets considered the
results for $\Delta M_2 / A_2$ show a rapid convergence toward $A_2$ for
values of $Q^2$ very close to the $Q_0^2$ values of Table 1. For higher
moments,  the convergence toward $A_n$ is still evident, but it is
slower probably because of the presence of resonances at large $\xi$. 

\indent The results presented in Figs. 5 and 6 correspond only to the 
contribution of the inelastic part of the structure functions to the
moments (\ref{1}). Since in Eq. (\ref{1}) the integration is extended up
to $\xi = 1$, the question of the role played by the contribution of the
elastic peak in the nucleon and the quasi-elastic peak in nuclei
naturally arises. Therefore, in case of the proton we have considered the
contribution $M_n^{(el)}(Q^2)$ resulting from the elastic peak, which
reads as
 \be
    M_n^{(el)}(Q^2) = {G_E^2(Q^2) + \tau ~ G_M^2(Q^2) \over 1 + \tau} ~
    {\xi_p^n \over 2 - \xi_p}
    \label{7bis}
 \ee
where $G_E$ ($G_M$) is the charge (magnetic) Sachs form factor of the 
proton, $\xi_p \equiv \xi(x = 1) = 2 / (1 + \sqrt{1 + 1 / \tau})$ and 
$\tau = Q^2 / 4 m^2$. In Fig. 7 we have reported the results obtained
for $M_n^{(tot)}(Q^2) \equiv M_n^{(el)}(Q^2) + M_n(Q^2)$ and $\Delta
M_n^{(tot)}(Q^2) / A_n(Q^2)$\footnote{We point out that Eq. (\ref{3})
holds only for $M_n^{(tot)}(Q^2)$ (cf. Ref. \cite{JU95}). The elastic
contribution (\ref{7bis}) {\em must} be included in Eq. (\ref{1}) when
the extraction of the higher-twists from the moments is required, which
is not the case of the present work.}. It can clearly be seen that the
higher-twists introduced by the proton elastic peak do not change
significantly (within few $\%$) the lowest-order moment $M_2(Q^2)$ for
$Q^2 \gsim 1.5 ~ (GeV/c)^2$, whereas they sharply affect higher moments
up to quite large values of $Q^2$ (cf. also Ref. \cite{JU95}). This means
that parton-hadron duality still holds for the total area
$M_2^{(tot)}(Q^2)$, i.e. for the average of the structure function over
all possible final states, with a mass scale consistent with the
one obtained including the inelastic channels only. On the contrary, at
least for $Q^2$ up to several $(GeV/c)^2$ the local duality is violated
by the elastic peak. This result is consistent with those of Refs.
\cite{BG70,RGP77}, where the applicability of the concept of
parton-hadron local duality in the region around the nucleon elastic
peak was found to be critical. We have obtained similar results in case
of the deuteron, while the analysis of the available nuclear data
appears to be compatible within large errors. To sum up, the twist-2
moments $A_n(Q^2)$ dominate the total moments $M_n^{(tot)}(Q^2)$
starting from values of $Q^2$ which strongly depend upon the order of
the moment (see Fig. 7). On the contrary, the $Q^2$ behaviour of the
inelastic contribution $M_n(Q^2)$ is governed by $A_n(Q^2)$ starting
from a value $Q^2 \simeq Q_0^2$ almost independent of the order of the
moment (see Figs. 5 and 6); thus, after Mellin transformation, the
twist-2 operators dominate the inelastic part of the structure function
for $Q^2 \gsim Q_0^2$ and, therefore, the parton-hadron local duality
holds for the averages of the structure function over the
nucleon-resonance bumps.

\indent Both the comparison of the moments and the discussion on the
onset of the local duality strongly suggest the occurrence of the
dominance of twist-2 operators in the inelastic structure functions for
values of $Q^2$ above the $Q_0^2$ values of Table 1. Therefore, we may
argue that present experimental data are compatible with a mass scale
$\mu^2 \simeq Q_0^2$; from Table 1 this means: $\mu_N^2 \simeq \mu_D^2 =
1.5 \pm 0.1 ~ (GeV/c)^2$ for the nucleon and the deuteron, and $\mu_A^2 =
1.0 \pm 0.2 ~ (GeV/c)^2$ for nuclei with $A \simeq 12$.

\indent A change of the mass scale of the twist-2 matrix elements in
nuclei ($\mu_A^2 < \mu_D^2 \simeq \mu_N^2$) is expected to lead to a
rescaling relation for the (inelastic) moments \cite{JA83,DT86}, viz.
 \be
    M_n^A(Q^2) = M_n^D(\delta_n(Q^2) \cdot Q^2)
    \label{8}
 \ee
with $\delta_n(Q^2 \simeq \mu^2) \simeq \mu_D^2 / \mu_A^2$. Assuming a
rescaling factor $\delta_n(\mu^2) \simeq \delta$ independent of the
order of the moment, the rescaling relation (\ref{8}) with $\delta =
1.17 \pm 0.09$ brings all the moments in the deuteron and in nuclei
into the best simultaneous agreement around the mass scale region, as
it is shown in Fig. 8. A possible mechanism for the $Q^2$-rescaling has
been suggested in Ref. \cite{JA83}: the quark confinement scale may
increase in going from a free nucleon to a nucleus, due to the partial
overlap of nucleons in the nuclear medium. The change in the quark
confinement size leads to a change in the mass scale $\mu^2$ and one
gets $\delta = \mu_D^2 / \mu_A^2 =$ $\lambda_A^2 / \lambda_D^2$, where
$\lambda_D$ ($\simeq \lambda_N$) and $\lambda_A$ are the average quark
confinement size in the deuteron (nucleon) and in the nucleus,
respectively. We stress that the partial quark deconfinement is not the
only mechanism yielding a rescaling effect; in this respect, we mention
also the model of Ref. \cite{DT86}, where the rescaling mechanism is
driven by the off-mass-shellness of the nucleon in the nucleus. 

\indent If $\delta_n(Q^2)$ is independent of the order of the moment,
then after Mellin transformation the $Q^2$- rescaling can be applied to
the nuclear structure function per nucleon $F_2^A(x, Q^2)$, viz.
 \be
    F_2^A(x, Q^2) = F_2^D(x, \delta(Q^2) \cdot Q^2)
    \label{9}
 \ee
where the virtual photon mass dependence of $\delta(Q^2)$ follows from
perturbative $QCD$ evolution at $NLO$ (see \cite{JA83,DT86}). In Fig. 9 
existing data on $F_2^A(x, Q^2)$ for nuclei with $A \simeq 12$ are
plotted for fixed values of $x$ in a wide $Q^2$ range. The {\em dual}
structure function (\ref{4}), properly folded \cite{SIM95} for taking
into account nuclear binding effects, is also shown in Fig. 9 for
different values of the mass scale ratio $\mu_D / \mu_A$. It can be
seen that at high $Q^2$ a mass scale ratio $\mu_D / \mu_A = 1.1$
removes most of the disagreement at large $x$, but at the price of
spoiling the agreement at intermediate values of $x$. Moreover, at low
values of $Q^2$ the present accuracy of the data does not allow any
serious discrimination between different quark confinement ratios. We
mention that in Ref. \cite{ZH92} a nucleon swelling corresponding to
$\mu_D / \mu_A \simeq 1.075$ has been derived from a combined analysis
of the $EMC$ effect at small and large $x$, performed within a
constituent-quark picture of the nucleon structure function in nuclei.

\indent To sum up, roughly consistent values of the mass scale ratio
$\mu_D / \mu_A$ between the deuteron and nuclei with $A \simeq 12$ can
be obtained in different ways, viz. 

\begin{itemize}

\item from the onset of local duality (see Table 1 and Fig. 5): $\mu_D /
\mu_A = 1.2 \pm 0.1$; 

\item from the rescaling of the moments at the static point $Q^2 \simeq
\mu^2$ (see Fig. 8): $\mu_D / \mu_A  = 1.08 \pm 0.05$;

\item from the $EMC$ effect (see Fig. 9): $\mu_D / \mu_A  \lsim 1.10$.

\end{itemize}

\indent Nucleon resonances contribute mostly to $M_4(Q^2)$ and
$M_6(Q^2)$ (see Fig. 6): the good overlap of these moments in the
deuteron and nuclei, observed in Fig. 8 after $Q^2$-rescaling, could
suggest that a change of the mass scale in nuclei might be consistent
with the more rapid decrease of the inelastic $P_{33}(1232)$ and
$D_{13}(1520)$ resonance form factors claimed in Ref. \cite{GE96},
where values of $Q^2$ as low as $0.1 ~ (GeV/c)^2$ are however involved.
The {\em experimental} suppression factor $R_s(Q^2)$, as determined in
Ref. \cite{GE96} using inclusive $^{12}C$ and $^{16}O$ data, is
reported in Fig. 10 as a function of $Q^2$ in the region of the
$P_{33}(1232)$ resonance production. Assuming that a constant
$Q^2$-rescaling $\delta(Q^2) \simeq \mu_D^2 / \mu_A^2$ can be used
below the static point ($Q^2 < \mu^2$), the suppression factor is
expected to be given by
 \be
    R_s(Q^2) = \left \{ {G_M^{(N - \Delta)}[{\mu_D^2 \over \mu_A^2} ~ 
    Q^2] \over G_M^{(N - \Delta)}(Q^2)} \right \}^2
    \label{10}
 \ee
where $G_M^{(N - \Delta)}(Q^2)$ is the magnetic form factor of the $N -
\Delta$ transition. Using a standard dipole form (for sake of
simplicity) and the value $\mu_D / \mu_A = 1.08$, one gets the solid
lines shown in Fig. 10 at $Q^2 \lsim 0.5 ~ (GeV/c)^2$. It can be seen
that (surprisingly) a simple $Q^2$-rescaling of the dipole ansatz is
consistent with the quenching observed for the $\Delta$-resonance
electroproduction both in case of the excitation strength alone (see
Fig. 10(a)) and for the total inclusive cross section (see Fig. 10(b) and
cf. Ref. \cite{CM93}).

\indent A $Q^2$-rescaling effect could in principle be applied also to
the elastic form factors of a nucleon bound in a nucleus and it can be
viewed as a change of the nucleon radius in the nuclear medium. In this
respect, it should be pointed out that: ~ i) in Ref. \cite{SI85} an
increase not larger than $\simeq 6 \%$ of the proton charge radius was
found to be compatible with $y$-scaling in $^3He$ and $^{56}Fe$; ~ ii)
the analysis of the Coulomb Sum Rule ($CSR$) made in Ref. \cite{CH91}
suggested an upper limit of $\simeq 10\%$ to the variation of the proton
charge radius in $^{56}Fe$; ~ iii) recently \cite{JO96} the experimental
value ($S_L$) of the $CSR$ in $^{12}C$ and $^{56}Fe$ has been accurately
analyzed at $Q^2 \simeq 0.3 ~ (GeV/c)^2$, obtaining a saturation value
$S_L = 0.94 \pm 0.13$ and $S_L = 0.97 \pm 0.12$, respectively. If at the
same value of $Q^2$ we would assume the same change of the mass scale
($\mu_D^2 / \mu_A^2 \simeq 1.15$) observed in the inelastic channels, an
increase of $\simeq 8 \%$ for the proton charge radius and a quenching
of $\simeq 0.84$ for $S_L$ would be obtained, both being at limit with
the quoted errors. However, based on the results shown in Figs. 5-7, it
is unlikely that a common $Q^2$-rescaling effect could be applied both to
the nucleon elastic peak and to the nucleon-resonance transitions.

\indent Before closing, let us make a brief comment on the
photoproduction of nucleon resonances. Real photon experiments
\cite{GE93} show that in several nuclei \cite{LNF93} a substantial
reduction of the excitation strength of $D_{13}(1520)$ and
$F_{15}(1680)$ resonances occur in comparison with the corresponding
hydrogen and deuterium data. This effect suggests both a broadening of
the resonance width and a quenching of their excitation strength
\cite{MSGR95}; while the broadening could be a consequence of final
state interactions \cite{AGL94}, the quenching might be ascribed to a
reduction of the transition strength in radial-type excitations, due
again to the overlap of confinement potentials among neighbouring
nucleons in nuclei \cite{GS94}. However, a common explanation of the
behaviour of the resonance bumps both in the photo- and in the
electro-production still awaits for a deeper understanding.

\section{Conclusions.}

\indent The concept of parton-hadron local duality represents a
very powerful tool for analyzing inclusive lepton scattering  data in
the low $Q^2$ region, where important quantities like the mass scale,
the leading and higher twists in the structure function may be
investigated. We have analyzed all the existing inclusive data on the
inelastic structure function of the nucleon, the deuteron and light
complex nuclei in a $Q^2$ range from $0.3$ to $5 ~ (GeV/c)^2$ and the
$Q^2$ behaviours of the structure function and its moments have been
presented for all the targets considered. In case of the proton we have
observed  that the Bloom-Gilman local duality is fulfilled only by the
inelastic part of the structure function, while the inclusion of the
contribution of the elastic peak leads to remarkable violations of the
local duality. In case of complex nuclei, despite the poor statistics of
the available data, our analysis suggests that the onset of the
parton-hadron local duality for the inelastic part of the structure
function is anticipated with respect to the case of the nucleon and the
deuteron. A possible interpretation of this result in terms of a
$Q^2$-rescaling effect has been discussed: using different methods, a
decrease of the mass scale of $\simeq 8 \%$ in nuclei turns out to be
consistent with inelastic experimental data. It has also been shown that
the same variation of the mass scale is consistent with the faster
fall-off of the $P_{33}(1232)$ transition form factors observed in
nuclei with respect to the nucleon case even at very low values of
$Q^2$. Finally, we expect that the same $Q^2$-rescaling effect cannot be
applied both to the elastic and transition form factors of a nucleon
bound in a nucleus, consistently with the severe constraints on
nucleon swelling arising from updated analyses of the Coulomb sum rule in
nuclei. 

\indent In conclusion, the Bloom-Gilman local duality in nucleon and
nuclei appears to be a non-trivial dynamical property of the inelastic
structure functions, whose deep understanding is still to be reached and
deserves much more attention from the theoretical as well as the
experimental point of view. As to the latter, more systematic and
high-precision inclusive data are needed for a clear-cut extraction of
information; our present analysis should therefore be considered as a
strong suggestion that interesting results can be expected, when the
structure functions of nucleon and nuclei are compared in the low $Q^2$
region. New facilities becoming operative in the next future, like
$CEBAF$, are expected to provide inclusive data with unprecedented
accuracy, allowing a throughout investigation of the relation among the
physics in the nucleon-resonance and Deep Inelastic Scattering regions.

\newpage

\vspace{4cm}

\small{\noindent Table 1. Values of $Q^2 = Q_0^2$ (in $GeV^2/c^2$) where a
systematic change in the slope is exhibited by the $Q^2$-dependence of
the moments $M_2(Q^2)$, $M_4(Q^2)$ and $M_6(Q^2)$ (see Fig. 5), computed
for the proton, the deuteron and nuclei with $A \simeq 12$ according to
Eq. (\ref{1}) using the experimental data shown in Figs. 1, 2 and 4.}

\vspace{1cm}

\begin{center}

\begin{tabular}{||c||c|c|c||}
\hline
             &     $M_2$     &     $M_4$     &     $M_6$     \\
\hline
$ proton   $ & $1.6 \pm 0.2$ & $1.5 \pm 0.1$ & $1.6 \pm 0.1$ \\
$ deuteron $ & $1.6 \pm 0.1$ & $1.5 \pm 0.1$ & $1.5 \pm 0.1$ \\
$ nuclei   $ & $0.8 \pm 0.1$ & $1.0 \pm 0.2$ & $1.2 \pm 0.1$ \\
\hline
\end{tabular}

\end{center}

\newpage

\begin{figure}

\epsfxsize=16cm \epsfig{file=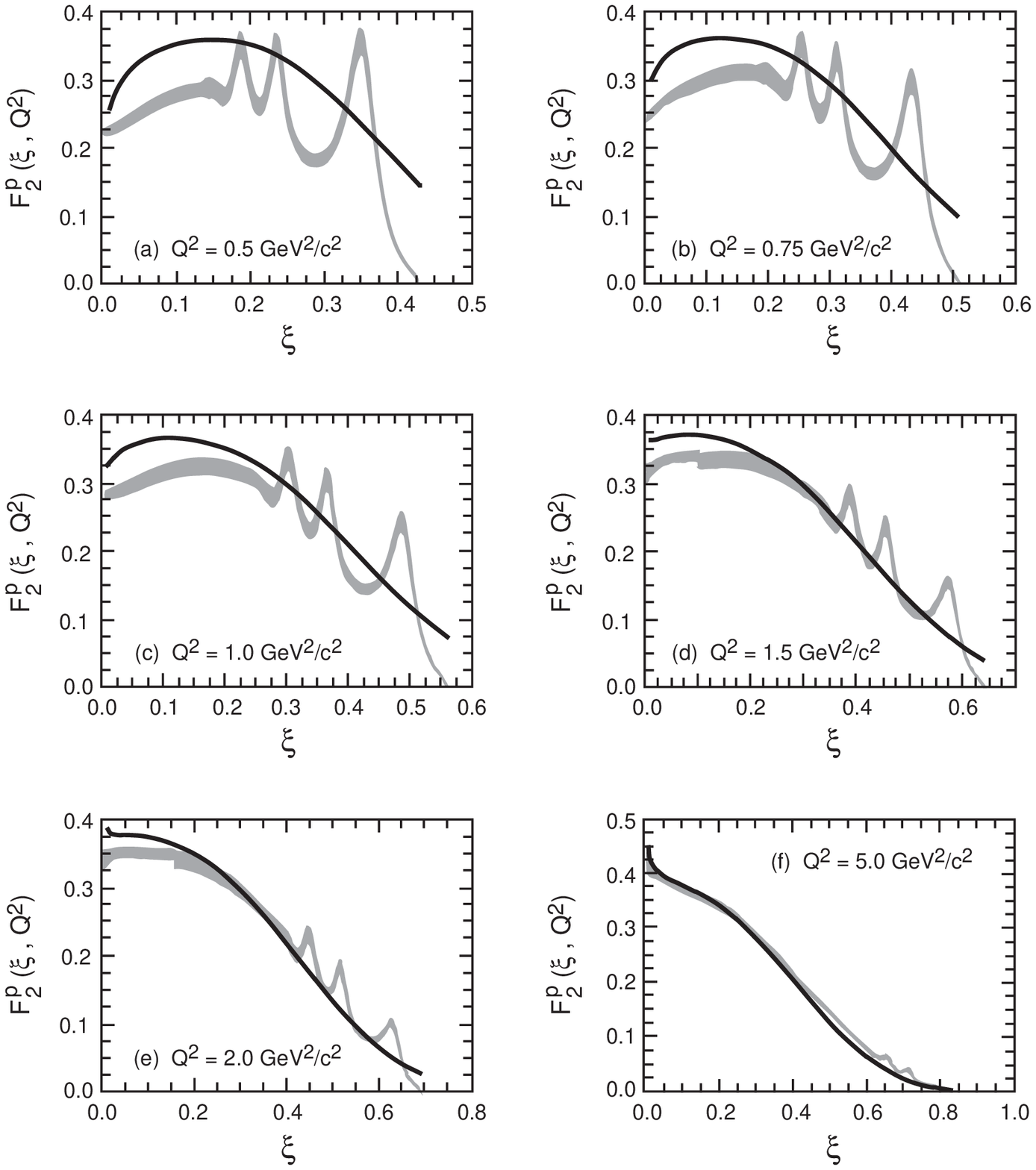}

\vspace{0.5cm}

\small{
\noindent Figure 1. Proton structure function $F_2^p(\xi, Q^2) = \nu
W_2^p(\xi, Q^2)$ versus the Natchmann variable $\xi$ (Eq. (\ref{2})) at
various values of $Q^2$. The shaded areas represent the pseudo-data,
obtained from a fit of $SLAC$ data at medium and large $x$ \cite{BO79}
and from a fit of $NMC$ data \cite{AR95} in the $DIS$ region, including
the total uncertainty of the fits. The solid lines are the {\em dual}
structure function of the nucleon (Eq. (\ref{4})), obtained starting
from the $GRV$ \cite{GRV92} parton densities evolved $NLO$ at low $Q^2$
and target-mass corrected according to Ref. \cite{GP76}.
}

\end{figure}

\newpage

\begin{figure}

\epsfxsize=16cm \epsfig{file=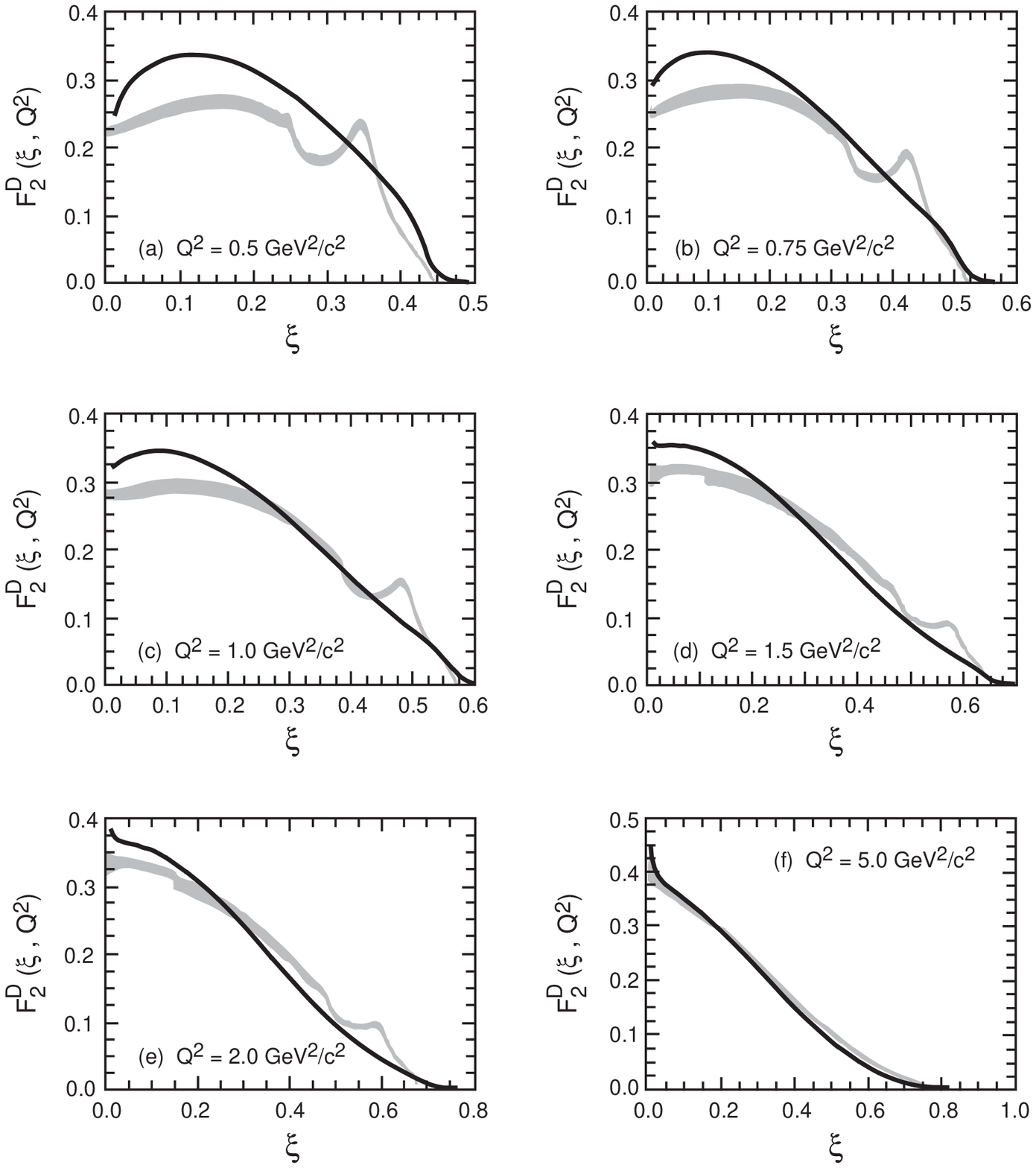}

\vspace{0.5cm}

\small{
\noindent Figure 2. The same as in Fig. 1, but for the deuteron
structure function per nucleon $F_2^D(\xi, Q^2) \equiv \nu W_2^D(\xi,
Q^2) / 2$. The $NMC$ data of Ref. \cite{AR90} at low $\xi$ and the fits
of inclusive data given in Refs. \cite{AR95} and \cite{BO79}, have been
considered. The solid curve is the {\em dual} structure function of the
nucleon (Eq. (\ref{4})), folded according to the procedure of Ref.
\cite{SIM95} with the nucleon momentum distribution in the deuteron
corresponding to the Paris nucleon-nucleon potential \cite{PARIS}.
}

\end{figure}

\newpage

\begin{figure}

\vspace{-1.5cm}

\epsfxsize=12cm \epsfig{file=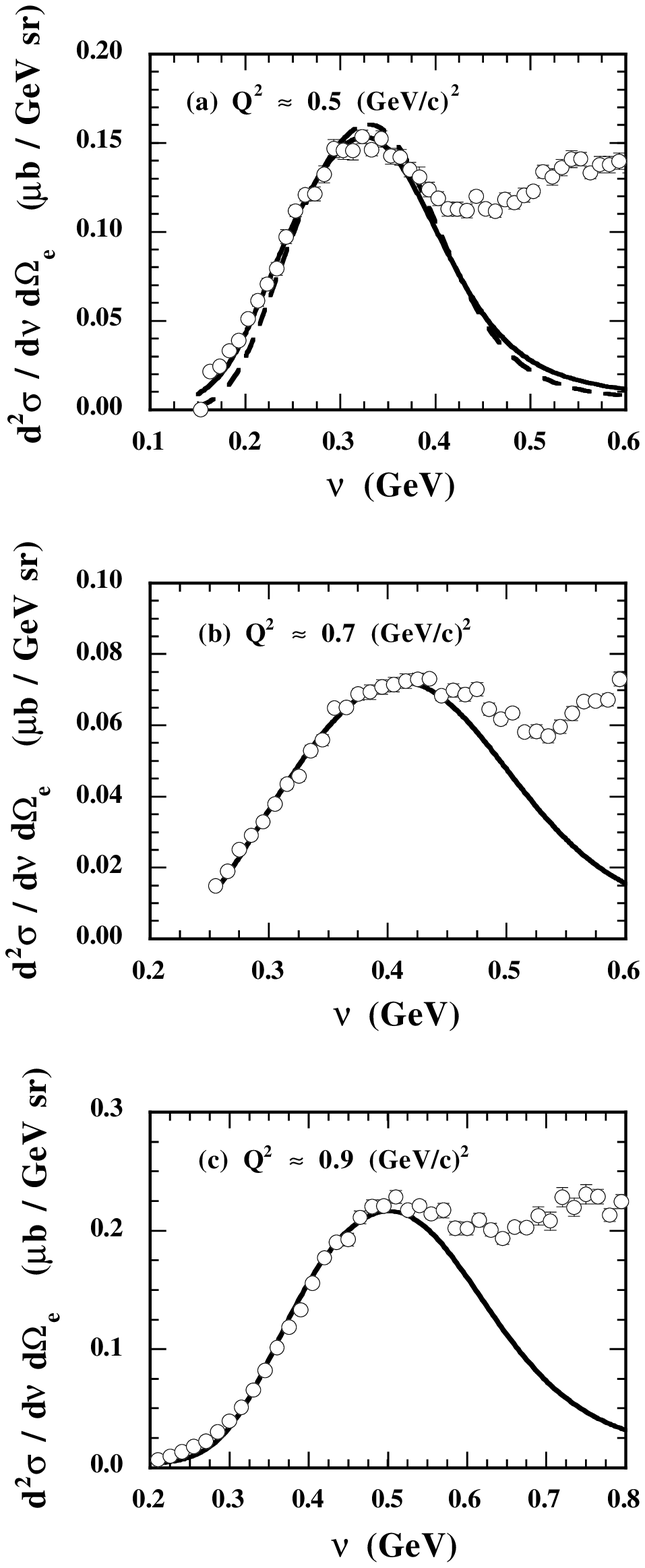}

\small{
\noindent Figure 3. Differential cross section (per nucleon) for the
inclusive process $^{12}C(e, e')X$ versus the energy transfer $\nu$ in
the $Q^2$ range from $0.5$ to $0.9 ~ (GeV/c)^2$ (at the quasi-elastic
peak). The solid lines represent the quasi-elastic contribution
calculated using the approach of Ref. \cite{CS96}, which includes final
state interaction effects. In (a) the dashed line is the same as the
solid line, but obtained within the impulse approximation only, i.e.
without including final state interaction effects. In (a), (b) and (c)
the electron beam energy and scattering angle, ($E_e, \theta_e$), are:
($1.3 ~ GeV, 37.5^o$), ($1.5 ~ GeV, 37.5^o$) and ($3.595 ~ GeV, 16^o$),
respectively. In (a) and (b) the experimental data are from Ref.
\cite{DA93}(c), while in (c) they are from Ref. \cite{DA93}(a).
}

\end{figure}

\newpage

\begin{figure}

\vspace{-4cm}

\epsfxsize=18cm \epsfig{file=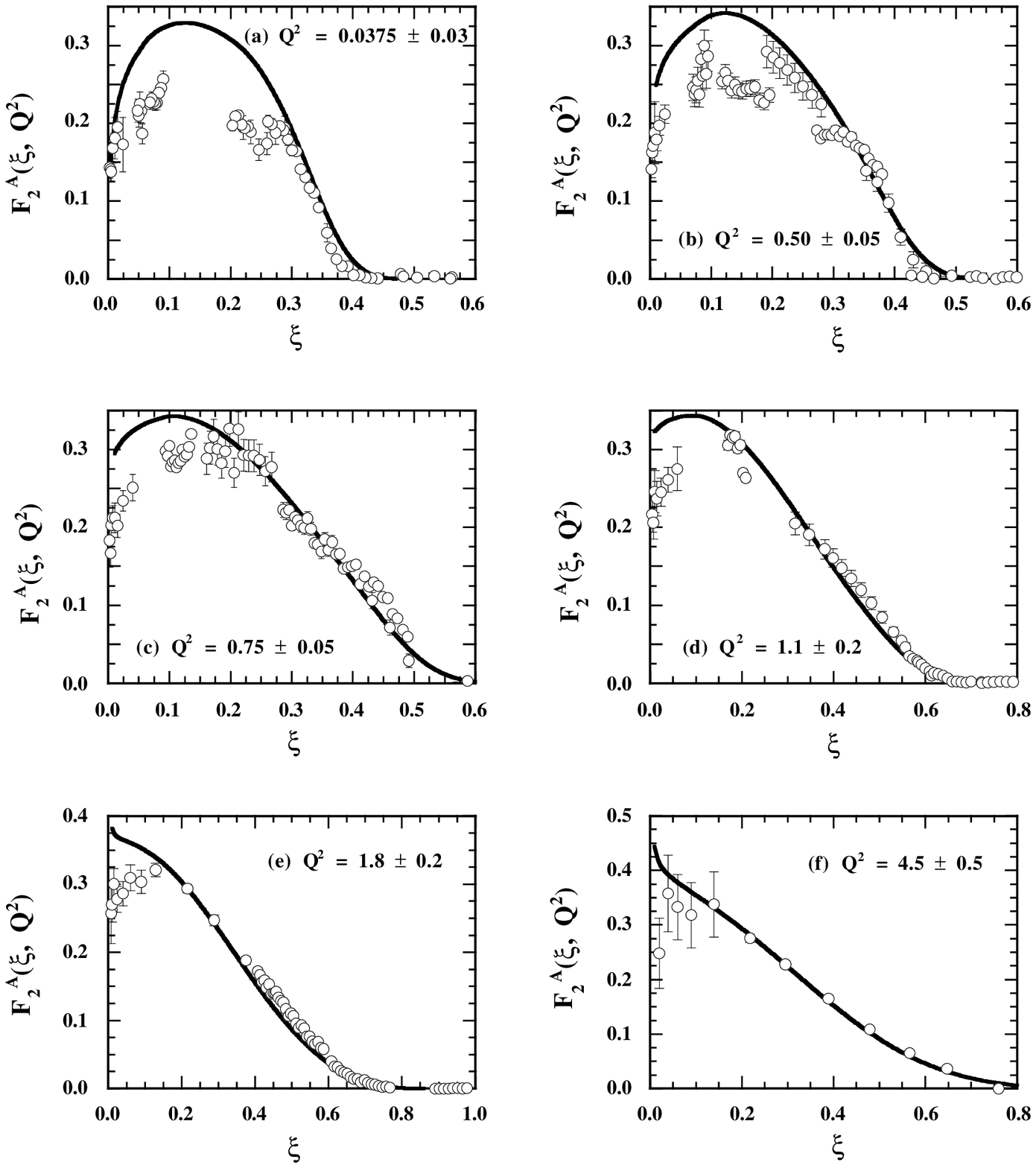}

\small{
\noindent Figure 4. The same as in Fig. 1, but for the nuclear structure
function per nucleon $F_2^A(\xi, Q^2) \equiv \nu W_2^A(\xi, Q^2) / A$
in case of nuclei with $A \simeq 12$. Experimental data are from Refs. 
\cite{GO94,FR81} ($^9Be$), \cite{BO79,AR90,DA93} ($^{12}C$) and
\cite{GE96} ($^{16}O$). The solid curve is the {\em dual} structure
function of the nucleon  (Eq. (\ref{4})), folded according to the
procedure of Ref. \cite{SIM95}, which adopts the nucleon spectral
function of Ref. \cite{CS96} to take into account nuclear binding
effects.
}

\end{figure}

\newpage

\begin{figure}

\vspace{-2cm}

\epsfxsize=12cm \epsfig{file=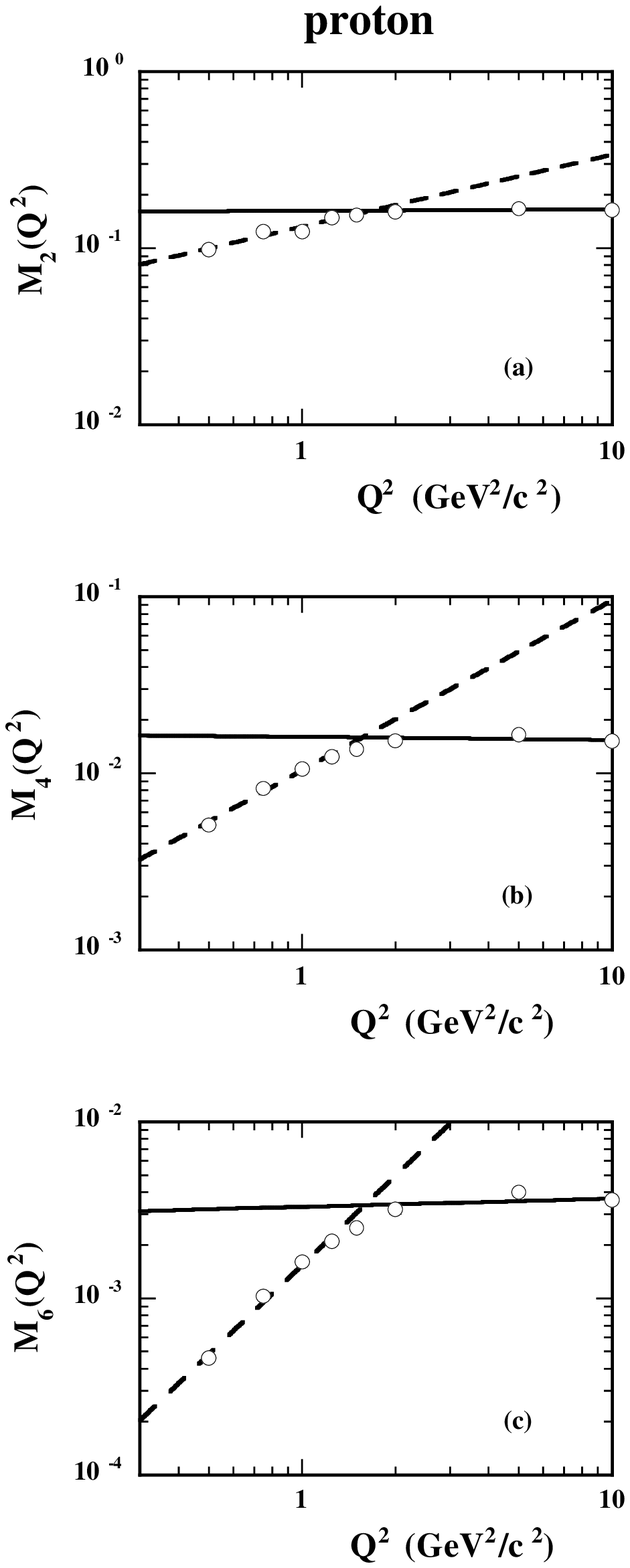}

\small{
\noindent Figure 5. The moments $M_2(Q^2)$, $M_4(Q^2)$ and $M_6(Q^2)$,
computed according to Eq. (\ref{1}) for the proton (a,b,c), the deuteron
(d,e,f) and nuclei with $A \simeq 12$ (g,h,i) using the experimental
data of Figs. 1, 2 and 4, versus $Q^2$. The dashed and solid lines are
linear fits to low and high $Q^2$ points and they are intended to show
up the change of the slope of the moments at $Q^2 \simeq Q_0^2$.
}

\end{figure}

\newpage

\begin{figure}

\vspace{-4cm}

\epsfxsize=12cm \epsfig{file=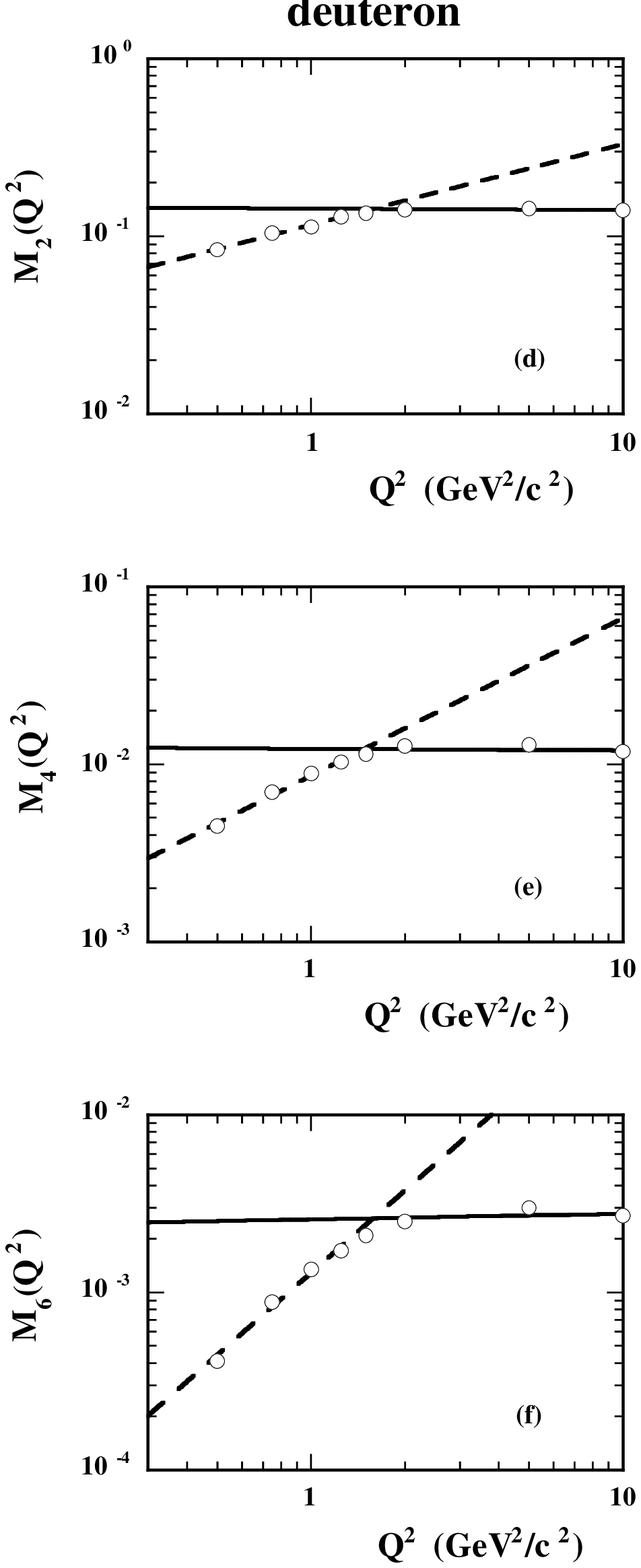}

\centerline{\small{\noindent Figure 5 - continued.}}

\end{figure}

\newpage

\begin{figure}

\vspace{-4cm}

\epsfxsize=12cm \epsfig{file=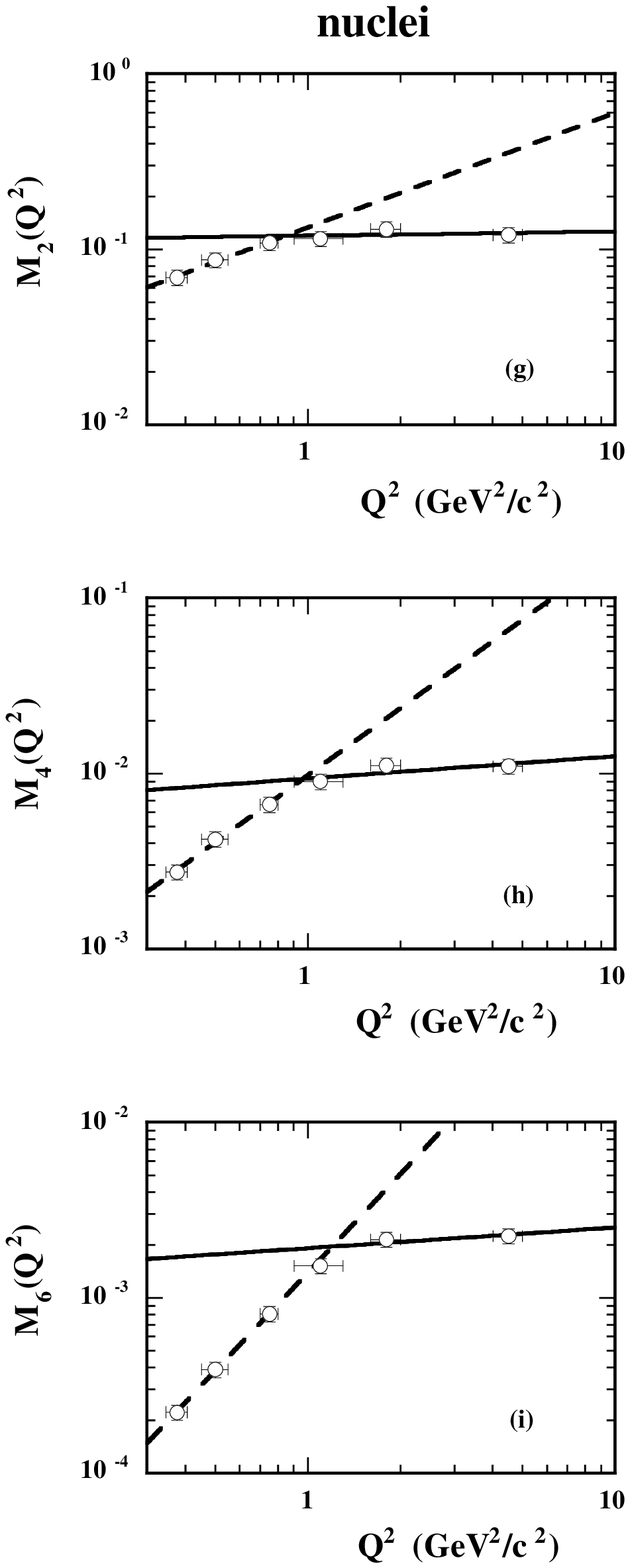}

\centerline{\small{\noindent Figure 5 - continued.}}

\end{figure}

\newpage

\begin{figure}

\vspace{-2cm}

\epsfxsize=12cm \epsfig{file=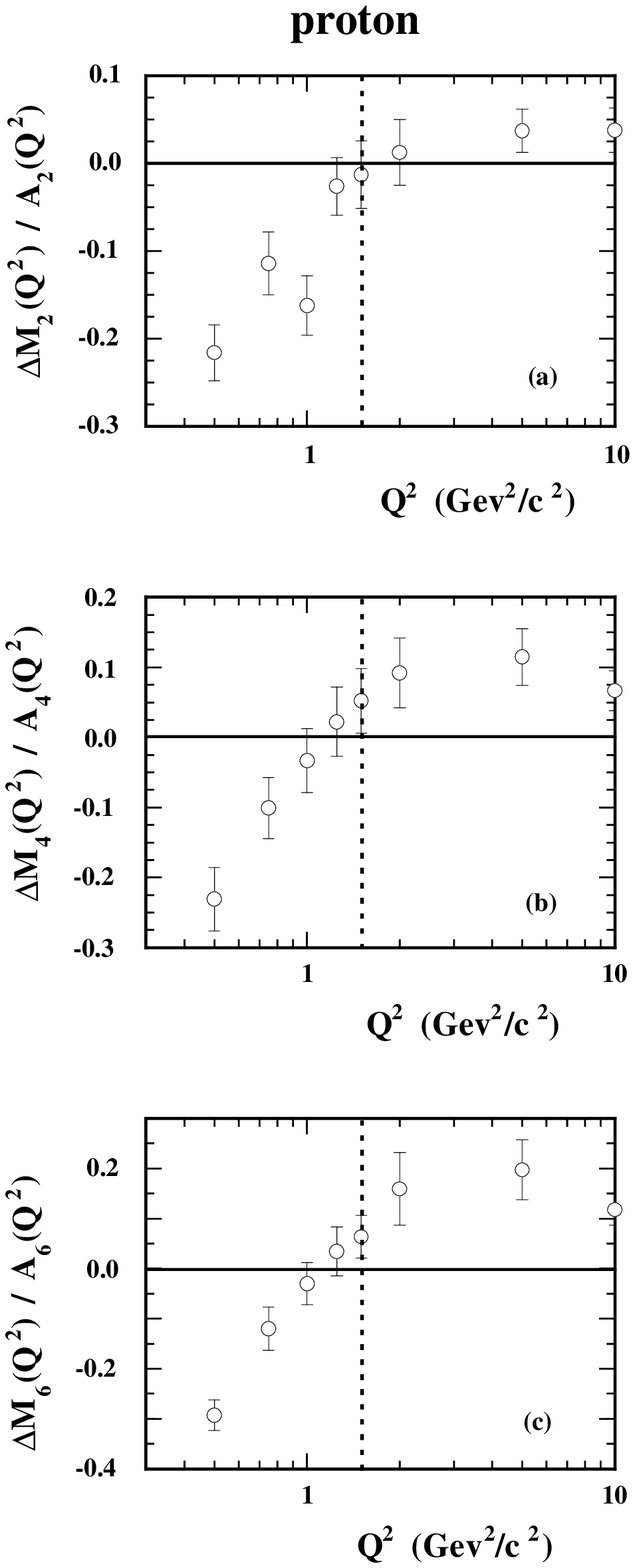}

\small{
\noindent Figure 6. The relative deviation $\Delta M_n(Q^2) / A_n(Q^2)$
(Eq. (\ref{7})) of the {\em experimental} moments $M_n(Q^2)$, reported
in Fig. 5, with respect to the leading twist moments $A_n(Q^2)$,
calculated from Eq. (\ref{1}) using the {\em dual} nucleon structure
function (\ref{4}), properly folded \cite{SIM95} with the nucleon
motion in the nuclear medium. The vertical dotted lines are intended
to show the approximate location of the static point $Q^2 \simeq
\mu^2$, namely $\mu_N^2 \simeq \mu_D^2 \simeq 1.5 ~ (GeV/c)^2$ and
$\mu_A^2 \simeq 1.0 ~ (GeV/c)^2$ (see text).
}

\end{figure}

\newpage

\begin{figure}

\vspace{-4cm}

\epsfxsize=12cm \epsfig{file=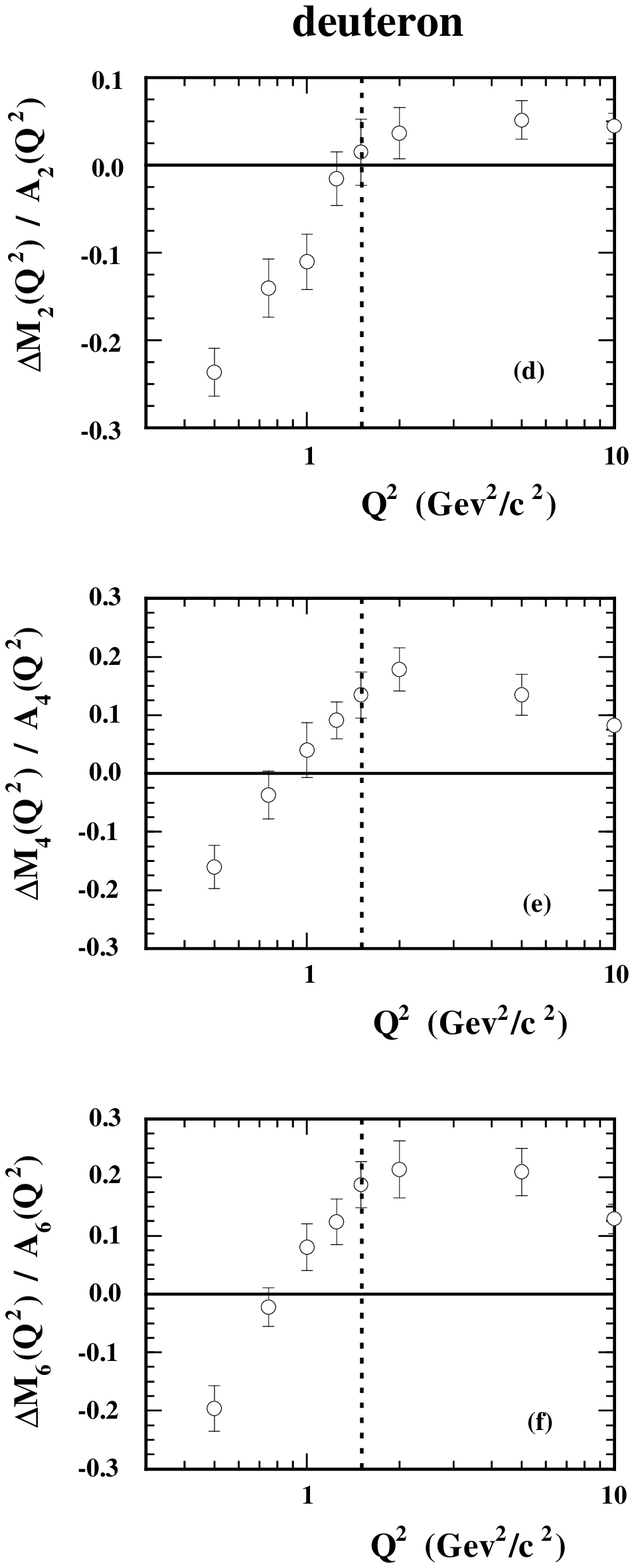}

\centerline{\small{\noindent Figure 6 - continued.}}

\end{figure}

\newpage

\begin{figure}

\vspace{-4cm}

\epsfxsize=12cm \epsfig{file=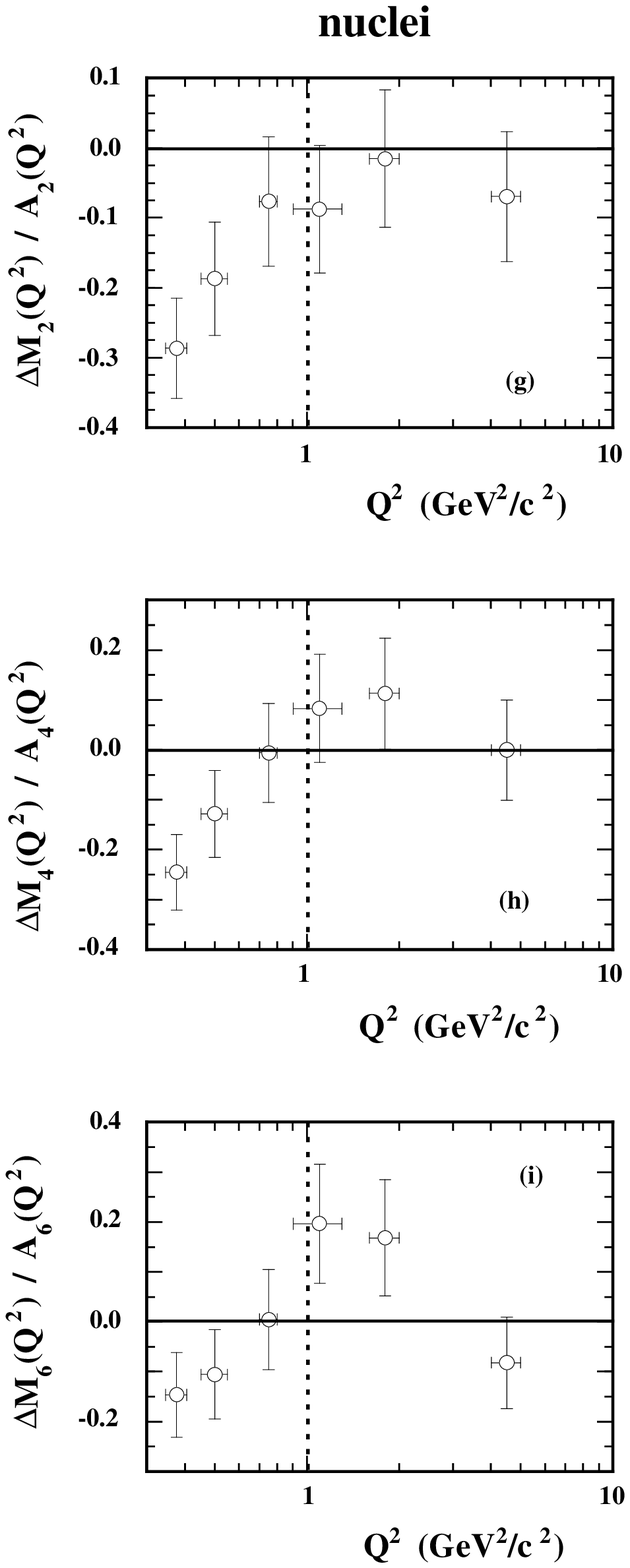}

\centerline{\small{\noindent Figure 6 - continued.}}

\end{figure}

\newpage

\begin{figure}

\vspace{-4cm}

\epsfxsize=18cm \epsfig{file=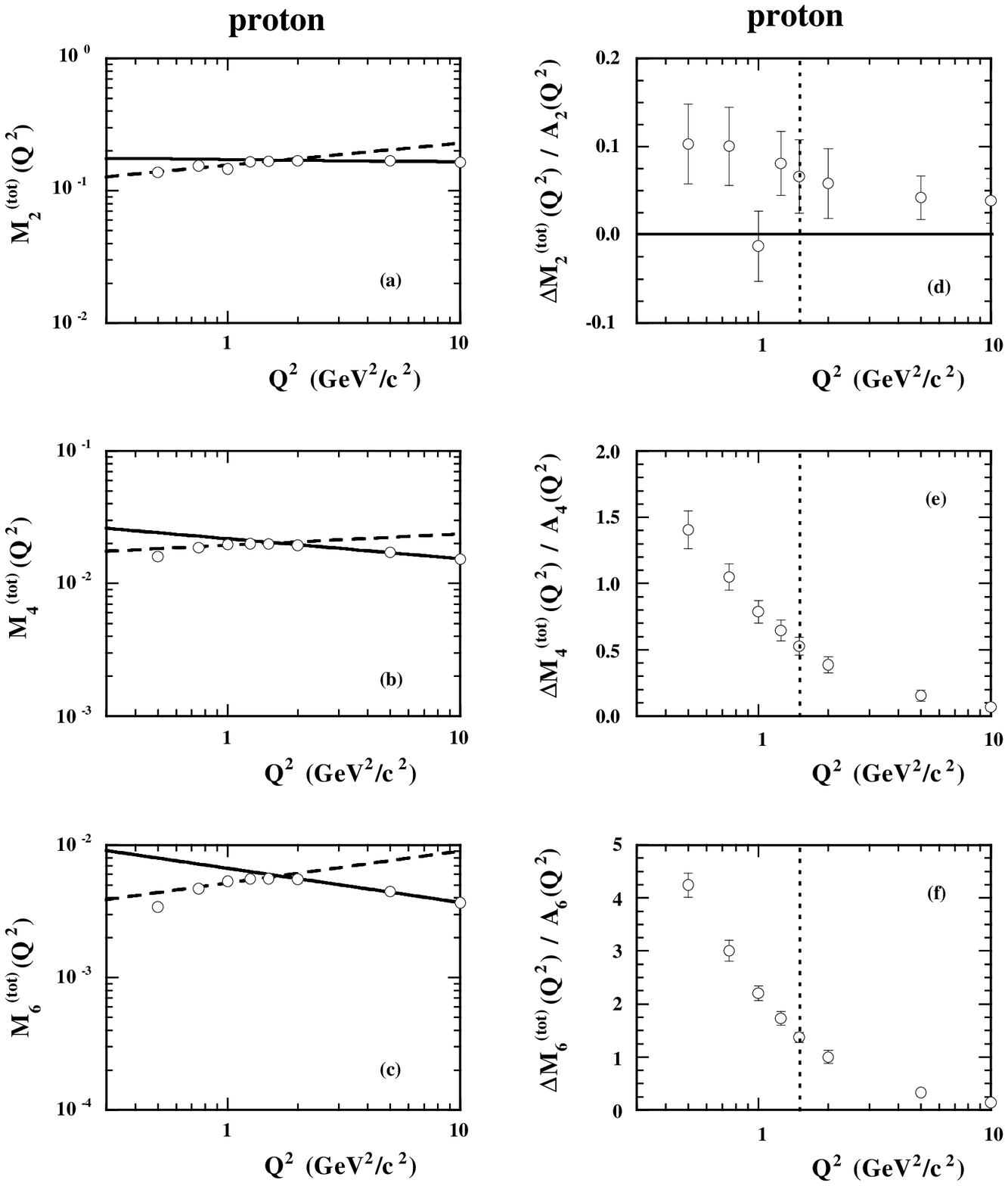}

\small{
\noindent Figure 7. The moments $M_n^{(tot)}(Q^2) = M_n^{(el)}(Q^2) +
M_n(Q^2)$ (a, b, c) and the relative  deviation $\Delta M_n^{(tot)}(Q^2)
/ A_n(Q^2)$ (d, e, f), calculated for the proton including in Eq.
(\ref{1}) the contribution of the elastic peak (\ref{7bis}), versus
$Q^2$. The vertical dotted lines are the same as in Fig. 6 (a, b, c).
}

\end{figure}

\newpage

\begin{figure}

\vspace{-3cm}

\epsfxsize=12cm \epsfig{file=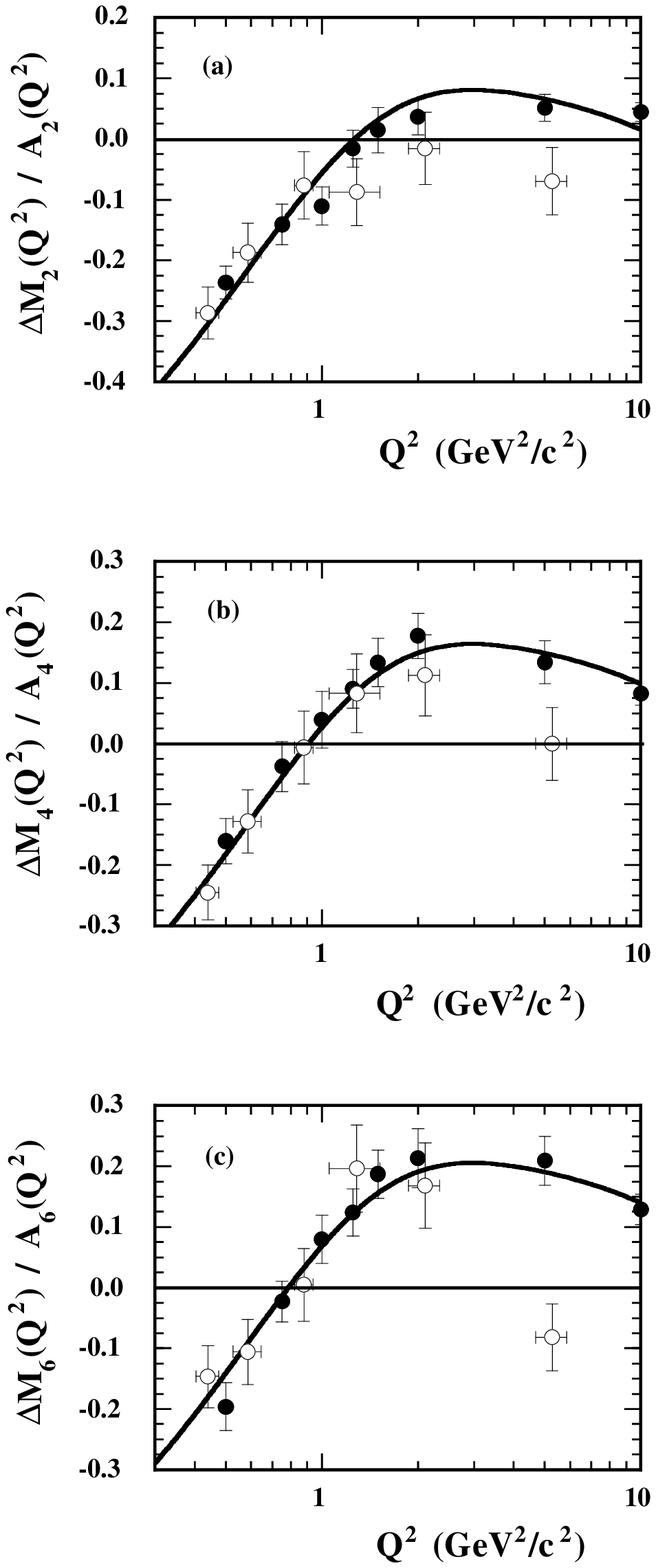}

\small{
\noindent Figure 8. The relative deviation $\Delta M_n(Q^2) / A_n(Q^2)$
(Eq. (\ref{7})) of the {\em experimental} moments $M_n(Q^2)$
calculated for the deuteron (black dots) and for nuclei with $A \simeq
12$ (open dots). The latter have been $Q^2$-rescaled according to Eq.
(\ref{8}) using $\delta_n(Q^2 \simeq \mu^2) \simeq \mu_D^2 / \mu_A^2 =
1.17$. The solid lines represent global fits of the deuteron and
($Q^2$-rescaled) nuclear points.
}

\end{figure}

\newpage

\begin{figure}

\vspace{-3cm}

\epsfxsize=16cm \epsfig{file=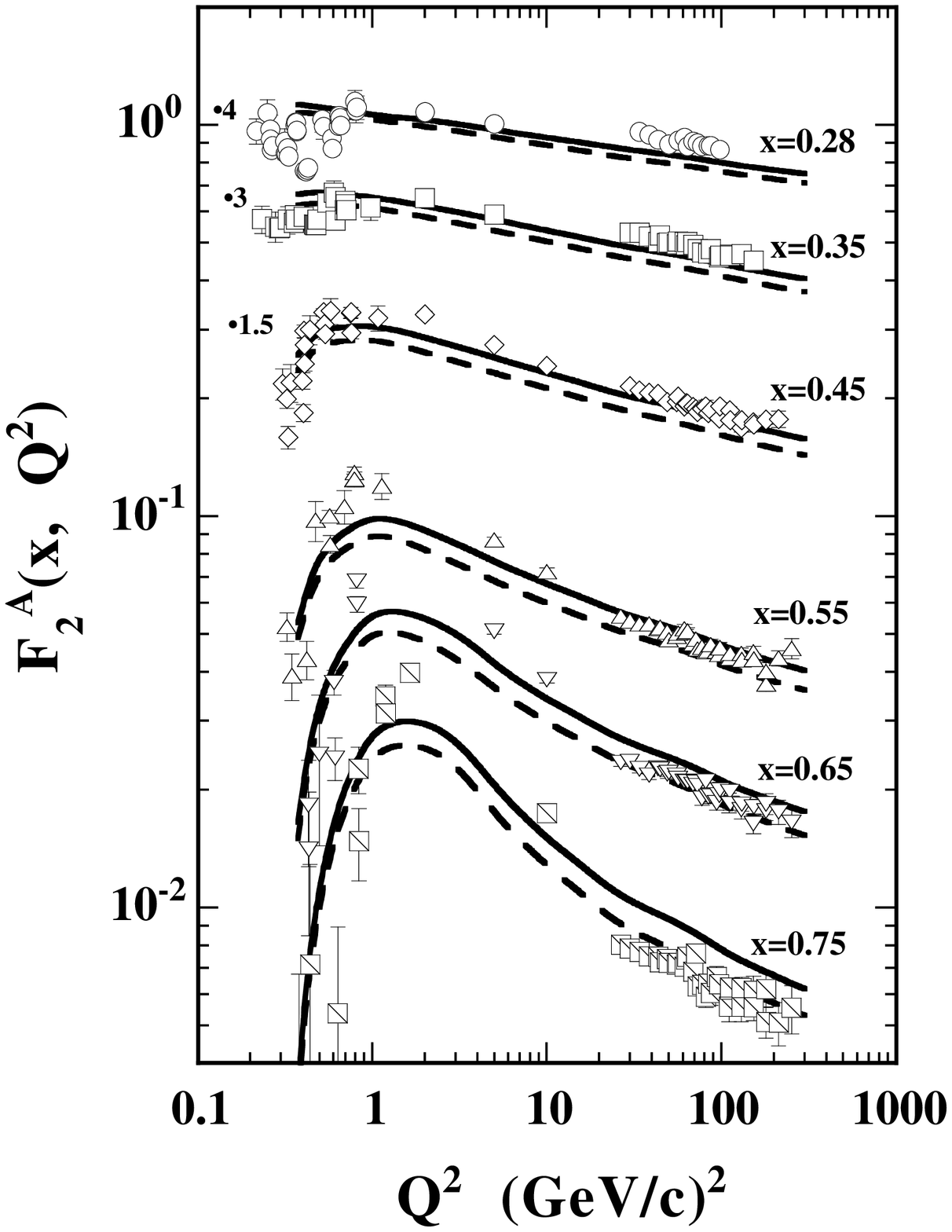}

\vspace{1cm}

\small{
\noindent Figure 9. Nuclear structure function per nucleon $F_2^A(x,
Q^2) \equiv \nu W_2^A(x, Q^2) / A$ versus $Q^2$ at fixed values of
$x$ in case of nuclei with $A \simeq 12$. The experimental data are from
Refs. \cite{GE96,BO79,AR90,GO94,FR81,DA93,BCDMS87}. The various
markers correspond to different values of $x$. The solid line is the
folding of the {\em dual} nucleon structure function (Eq. (\ref{4}))
obtained using the procedure of Ref. \cite{SIM95}. The dashed line is
the same as the solid line, but using the rescaling relation (\ref{9})
with $\delta(Q^2)$ taken from Ref. \cite{JA83} and $\mu_D / \mu_A =
1.10$.
}

\end{figure}

\newpage

\begin{figure}

\vspace{-2cm}

\epsfxsize=15cm \epsfig{file=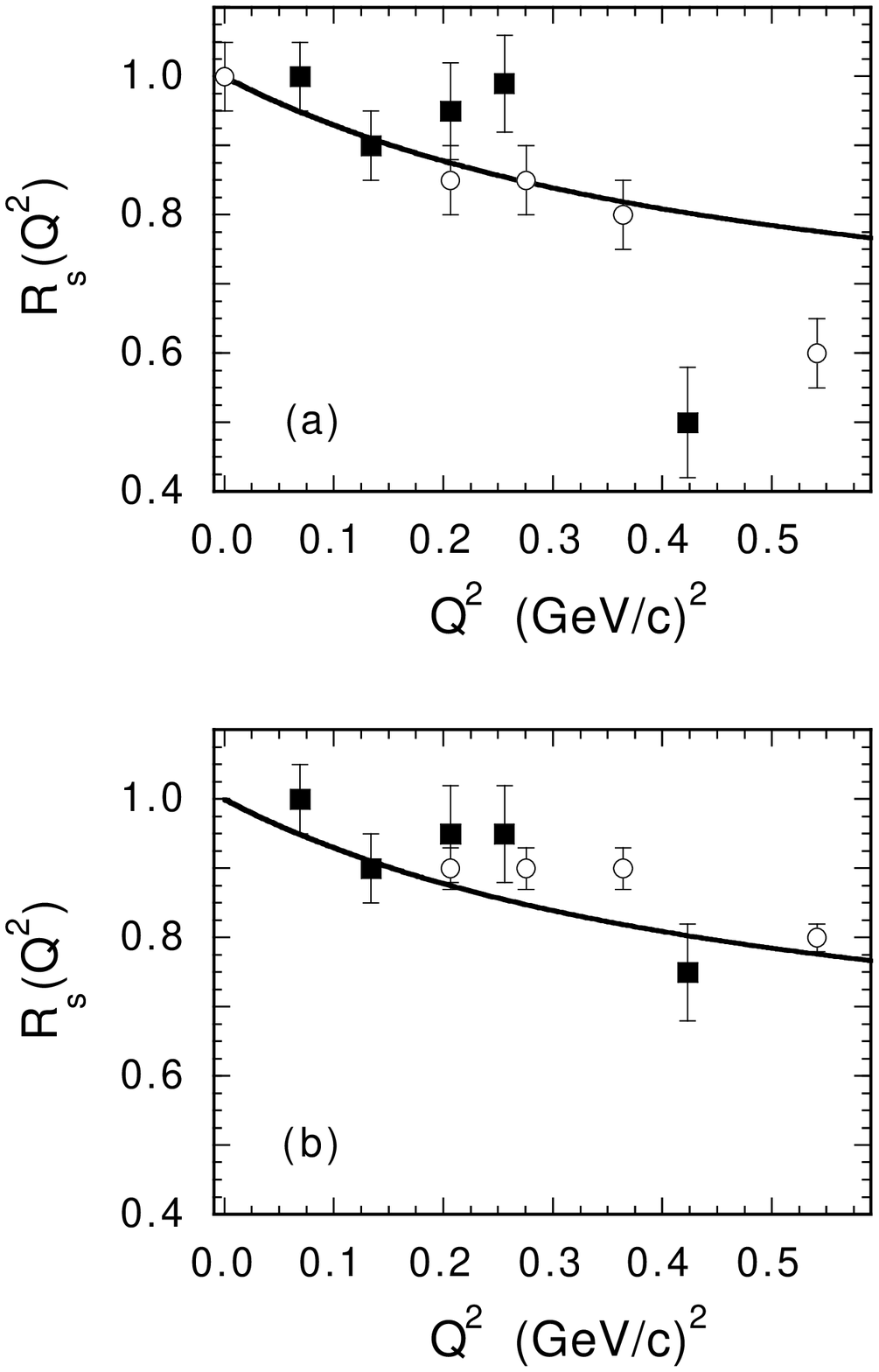}

\vspace{0.5cm}

\small{
\noindent Figure 10. The suppression factor $R_s(Q^2)$ versus $Q^2$, as
determined in Ref. \cite{GE96}, using inclusive $^{12}C$ (open dots)
and $^{16}O$ data (full squares), in case of the excitation of the
$P_{33}(1232)$ resonance only (a) and for the total inclusive cross
section $@ ~ W = 1232 ~ MeV$ (b). The solid line is the prediction  of
the rescaling relation (\ref{10}), obtained adopting a dipole ansatz for
the magnetic form factor of the $N - \Delta$ transition, $G_M^{(N -
\Delta)}(Q^2)$, and the value $\mu_D / \mu_A = 1.08$.
}

\end{figure}

\end{document}